\documentclass[aps,pre,twocolumn,superscriptaddress,showpacs,showkeys]{revtex4}

\usepackage{natbib}
\usepackage{dcolumn}
\usepackage{amsmath}
\usepackage{graphicx}
\usepackage{rotating}
\usepackage{multirow}

\begin{document}
\title{Schr\"{o}edinger-like PageRank equation and localization in the WWW}

\author{Nicola Perra}
\affiliation{Dep of Physics, SLACS-CNR University of Cagliari, S.P. Monserrato-Sestu Km 0.700, 09042 Cagliari Italy}
\affiliation{Linkalab, Complex Systems Computational Lab. 09100 Cagliari Italy}

\author{Vinko Zlatic}
\affiliation{Centre SMC CNR-INFM, Dep. Physics University ``Sapienza'' P.le Moro 5 00185 Rome, Italy}
\affiliation{Theoretical Physics Division, Rudjer Boskovic Institute, P.O.Box 180, HR-10002 Zagreb, Croatia}

\author{Alessandro Chessa}
\affiliation{Dep of Physics, SLACS-CNR University of Cagliari, S.P. Monserrato-Sestu Km 0.700, 09042 Cagliari Italy}
\affiliation{Linkalab, Complex Systems Computational Lab. 09100 Cagliari Italy}

\author{Claudio Conti}
\affiliation{Centre SOFT CNR-INFM, Dep. Physics University ``Sapienza'' P.le Moro 5 00185 Rome, Italy}

\author{Debora Donato}
\affiliation{Yahoo! Research  C. Ocata 1, 08003 Barcelona, Spain.}

\author{Guido Caldarelli}
\affiliation{Centre SMC CNR-INFM, Dep. Physics University ``Sapienza'' P.le Moro 5 00185 Rome, Italy}
\affiliation{Linkalab, Complex Systems Computational Lab. 09100 Cagliari Italy}

\date{\today} \widetext

\begin{abstract}
The World Wide Web is one of the most important communication systems we use in 
our everyday life. 
Despite its central role, the growth and the development of the WWW is 
not controlled by any central authority. 
This situation has created a huge ensemble of connections whose complexity 
can be fruitfully described and 
quantified by network theory~\cite{1_08,2_08,3_08,4_08}. 
One important application to sort out the information present in these connections 
is given by the PageRank alghorithm\cite{5_08}.
Computation of this quantity is usually made iteratively with a large use of 
computational time.
In this paper we show that the PageRank can be expressed in terms of a wave function 
obeying a 
Schr\"{o}dinger-like equation. In particular the topological disorder given 
by the unbalance of outgoing 
and ingoing links between pages, induces wave function and potential structuring. 
This allows to directly 
localize the pages with the largest score. Through this new representation 
we can now compute the PageRank 
without iterative techniques. For most of the cases of interest 
our method is faster than the original one. 
Our results also clarify the role of topology in the diffusion of 
information within complex networks. 
The whole approach opens the possibility to novel techniques inspired by quantum physics for the analysis of 
the WWW properties. 
\end{abstract}
\maketitle       
Most of the work in the field of complex networks has been related to the analysis of 
particular datasets and their modelling through suitably chosen statistical growth models~\cite{1_08,2_08,6_08}. 
A more recent approach of particular interest is the study of the physical processes 
taking place on a network and interacting with its topology~\cite{7_08,8_08} as, 
for example, the diffusion on a complex network. 
The large variability in the number of neighbours that characterises complex 
networks produces unexpected behaviour in the propagation of viruses and diseases~\cite{9_08}. 
The same topological randomness makes also unpredictable the behaviour of the 
PageRank~\cite{5_08}. 
 
\begin{figure}[t]
    \includegraphics[width=8cm]{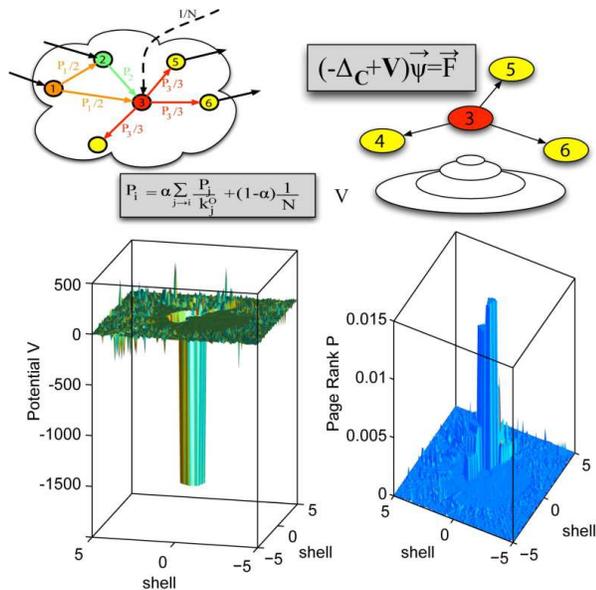}
    \caption{\label{fig1} Top a pictorial representation of the Page Rank equation and its counterpart in term of 
wave function. Bottom 3-d plot of potential $V$ and the corresponding PageRank measured along concentric shells 
around an original vertex.}
\end{figure}

\begin{figure}[t]
    \includegraphics[width=8cm]{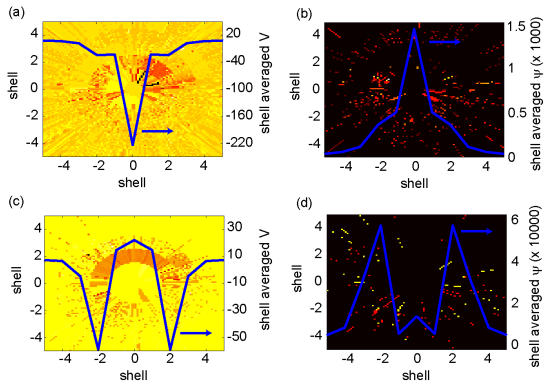}
    \caption{\label{fig2} Contour plot of potential and wave function for two opposite cases of attractive (a,b) 
and repulsive (c,d) potential. As before they are computed over concentric shells of neighbours computed through a BFS 
algorithm. In both cases the peak of the potential are produced by the topological disorder given by the difference 
between in-degree and out-degree of a node. The continuous line is obtained by averaging over nodes on the same shell. }
\end{figure}

\begin{figure}[t]
    \includegraphics[width=8cm]{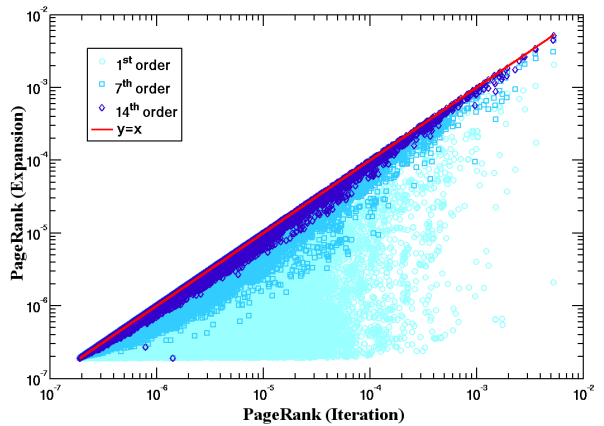}
    \caption{\label{fig3} A plot of the convergence of the expansion for the PageRank. On the x-axis we have the 
PageRank computed in the traditional iterative way, on the y-axis the PageRank computed through our proposed expansion. 
Every vertex corresponds to a different page. Increasing the order of the expansion the points condensate on the curve 
$y=x$. This process occurs very quickly for the top $100$ ranking sites of the WWW portion we studied.}
\end{figure}

This quantity (firstly introduced by the Google search engine founders) measures the importance of a page 
in the entire system~\cite{5_08} (see Appendix). The present procedure to compute this quantity on a given graph is based on numerical iterations. The whole process corresponds to surf the WWW with a number of random surfers and to assign the PageRank of a page as the number of times it is crossed by a walker. Just to avoid the presence of trapping 
states for these surfers, they are allowed to jump into a completely new page of the WWW (with probability 
$(1-\alpha)\sim 0.15$). The role of the damping factor $\alpha$ has been intensively studied, since PageRank 
changes significantly when $\alpha$ is modified~\cite{10_08}.\\
The effectiveness of this ranking procedure is witnessed by the success of the search engine Google. 
Since the first investigations, various procedures and algorithms have been presented to determine the PageRank 
and its dynamical evolution~\cite{11_08,12_08,13_08,17_08}. It is fair to say that the study of PageRank constitutes a 
field of research on its own. Actually the PageRank has also a strong impact in may Web-related topics such as 
propagation of trust (e.g. TrustRank, propagation of trust and distrust), and smoothed classification of pages.\\
We present here a novel approach based on an analogy with quantum physics. In particular we show that it is possible 
to rearrange the terms of the PageRank equation in such a way as to obtain a Schr\"{o}dinger-like equation for a wave
 function. For a given page, the wave function is given by the PageRank divided by the out-degree (i.e., the number of 
outgoing links). Through this new approach we recognize that the topology of the system, (more precisely the 
difference between the number of ingoing and outgoing links per page), plays the role of a local quantum-like 
potential $V$ defined on every page.  Two basic assumptions underlie these results. Firstly, in the Schr\"{o}dinger 
equation for the WWW the spatial derivatives in the Laplacian operator are expressed in the discrete space of a graph 
\cite{14_08}. Secondly, we adopt a directed version of this operator to account for the real topological structure 
of the Web (see Appendix for a discussion on this point).\\
While the first operator in the new equation can be considered a Laplacian operator, the second operator $V$ plays 
the role of a local potential (given by the out-degree of a page divided by the parameter $\alpha$ minus the 
in-degree of the same page). This is a crucial point, since through the use of this potential, we can distinguish 
between the case of complete reciprocity~\cite{15_08} between nodes and the directed case. In the former case we 
show that the wave function must be constant everywhere. When reciprocity is broken, we have that a non trivial 
PageRank distribution is attained. A final constant term is also present and it is given by $(1-\alpha)/\alpha N$ 
where $N$ is the total number of pages in the system. With respect to the standard Schr\"{o}dinger equation, this 
constant term describe the behaviour of a source. Every page creates a minimum quantity of PageRank that accumulates 
in most linked (successful) pages. Even before considering the particular cases in which the equation can be 
analytically solved it is possible to understand the ``physical'' meaning of these various terms. The discrete 
Laplacian directed operator accounts for the diffusive nature of the PageRank, and the potential accounts for the 
effects of the topology in the diffusion process. The constant term that vanishes when $\alpha$ tends to $1$ has the 
form of a source of charge and it is related to the possibility that after a large jump, a walker can be originated 
in the page. In Figure \ref{fig1} we show the potential $V$ and the corresponding PageRank measured along concentric 
shells around an original vertex.\\
Following this physical analogy, some limit situations can be immediately understood. Consider the case in which 
the network  is undirected, this means that the in-degree of every page is equal to the out-degree. No trapping 
states are present and we can therefore consider $\alpha=1$. The Schr\"{o}dinger equation becomes then a Laplace 
equation whose solution is given by a constant wave function. We recover then the result that the PageRank in this 
case is proportional to the degree of the page. This clarifies the role of the potential: it is the disorder given 
by the absence of reciprocity in the links that creates the non trivial properties of the PageRank observed in the 
WWW. When $V$ is different than $0$ and displays a minimum, (as in the case of an electron wave function in a 
disordered medium) the wave function is localized into the well of such a potential. In Figure \ref{fig2} we show 
some examples of the topology of the potential in proximity of some hubs and the corresponding wave-functions; the 
wave function is highly peaked in correspondence of the minima of $V$. Similarly, those nodes that have a low PageRank 
tend to be located in proximity of the maximum of the potential. It is important to stress that, since the Web has not 
a simple topology, the fact that a page is minimum or a maximum of the potential is only evident when plotting the values 
at the nearest neighbours in the network. By exploiting the topology information in the potential $V$, 
one is able to gain information about the PageRank spatial distribution. 
In particular, according to our picture, the higher scores of the PageRank will condensate within 
the potential wells. As a very first approximation one can consider 
pages corresponding to the smallest values of $V$. A simple analysis on the a subset of the WWW 
(the 2005 .eu domain~\cite{16_08}) shows that by considering the $100$ minimal values of the potential 
$V$ one can guess (irrespective of the order) 61 among the top 100 values of PageRank.
To improve this result (without considering the general solution for $\psi$), one would have to take into account gradients and expansions around the local $V$ minima. 
Actually our interpretation suggests a relatively simple way to compute the whole distribution of the PageRank 
(independently from the physical interpretation of the directed Laplacian operator). 
In principle once the matrices of the Laplacian operator and the potential operator are known, 
the wave function (and henceforth the set of PageRank values) could be computed by inverting these operators. 
This simple operation is unfeasible when the size of the matrix is given by the tens of billions of pages composing 
the WWW. Here instead we adopt a different approach based on matrix expansion that in principle could allow also to 
study the time evolution. The result of this matrix expansion are plotted against the value of PageRank (computed in 
the traditional way onto the same subset of the WWW cited above) is shown in Figure \ref{fig3}.  One can increase as 
desired the order of the expansion with a computational cost that increases only linearly with the order (see Appendix 
for details).\\
In conclusion, rearranging the terms in the PageRank equation, we have been able to obtain a Schr\"{o}dinger-like 
equation allowing us to define a site potential $V$ in the system. According to the meaning of the original equation, 
the PageRank (now in the form of  a wave function), localizes in the minima of this potential allowing educated 
guesses on the distribution of PageRank score. Furthermore, this representation makes  possible to compute the value 
of PageRank score through matrix expansion rather than iteratively. While in the limit of infinite pages the latter
 method is certainly less time consuming, for the subset of the WWW analysed here the expansion method is noticeably 
faster.  Since the PageRank is also used for a variety of Web-related problems (like for example the propagation of 
trust), in most of these cases where one deals with systems of medium-large size, our approach can be competitive. 
Finally, the formal analogy with quantum physics makes available a complete series of theoretical frameworks like 
perturbation theory that could be used to study the dynamical evolution of the PageRank score distribution.  
While with iterative techniques we must compute anew the PageRank score distribution for any change of the topology, 
through our approach we can in principle obtain in a much shorter time the evolution of this quantity.
The Schr\"{o}dinger-like equation we introduce, formally generates a connection among completely different kind of 
phenomena and at very distant scales, like the World Wide Web and the microscopic world of Quantum Physics and it 
is likely to pave the way to a series of activities in this field.

\section{Appendix}

\subsection{PageRank and wave equation}

The equation for the PageRank $P_{i}$ of a page $i$ is given by:
\begin{equation}
\label{1_ap}
P_{i}=\alpha\sum_{j \rightarrow i}\frac{P_{j}}{k^{O}_{j}}+(1-\alpha)\frac{1}{N}
\end{equation}
where $\alpha$ is a parameter tuned to $0.85$ in the first application made on WWW and $N$ is the total number of 
pages in the system. The quantity $k^{O}_{j}$ indicates the out-degree of page $j$. We can modify the above equation 
by introducing the quantity
\begin{equation}
\psi_{i} =\frac{P_{i}}{k^{O}_{j}}
\end{equation}
defined everywhere apart the dangling nodes for which  $k^{O}_{j}=0$.\\ 
The above equation then becomes
\begin{equation}
 \psi_{i}=\frac{\alpha}{k^{O}_{i}}\sum_{j \rightarrow i}\psi_{j}+\frac{(1-\alpha)}{k^{O}_{i}}\frac{1}{N}
\end{equation}
With a little algebra we obtain
\begin{equation}
\psi_{i}=\frac{\alpha}{k^{O}_{i}}\left[ \sum_{j \rightarrow i}\psi_{j}-k^{I}_{i}\psi_{i}\right]+\alpha \frac{k^{I}_{i}}{k^{O}_{i}}\psi_{i}+ \frac{(1-\alpha)}{k^{O}_{i}}\frac{1}{N} 
\end{equation}
where he quantity $k^{I}_{i}$ indicates the in-degree of page $i$.
Rearranging the terms with $\psi_{i}$ we obtain
\begin{equation}
 -\frac{\alpha}{k^{O}_{i}}\left[ \sum_{j \rightarrow i}\psi_{j}-k^{I}_{i}\psi_{i} \right]+\left( 1-\alpha\frac{k^{I}_{i}}{k^{O}_{i}}\right) \psi_{i}=\frac{(1-\alpha)}{k^{O}_{i}}\frac{1}{N} 
\end{equation}
multiplying both sides for $k^{O}_{i}/\alpha$ we finally obtain
\begin{equation}
\label{6_ap}
 -\Delta_{C}(\psi_{i})+V_{i}\psi_{i}=\frac{(1-\alpha)}{\alpha}\frac{1}{N}
\end{equation}
where we have defined
\begin{equation}
 \Delta_{C}(\psi_{i})=\sum_{j \rightarrow i}\psi_{j}-k^{I}_{i}\psi_{i} \,\ \,\ and \,\ \,\ V_{i}=\frac{k^{O}_{i}-\alpha k^{I}_{i}}{\alpha}
\end{equation}
The function  $\Delta_{C}$ is the directed counterpart of the action of the Laplacian operator $\Delta$ acting on 
a vector whose $N$ elements are the $\psi_{i}$. Passing to the vectorial representation eq. (\ref{6_ap}) becomes 
\begin{equation}
\label{8_ap}
 \left( -\Delta_{C}+V\right){\psi}={F} 
\end{equation}
where $V$ is a diagonal matrix whose elements are given by the $V_{i}$ and whose effect on the vector is to multiply 
the element $\psi_{i}$ with $V_{i}$ and $F$ is a vector whose all elements have the value $(1-\alpha)/(\alpha N)$.
\subsection{Role of Directed Laplacian Operator}

\begin{figure}[t]
    \includegraphics[width=9cm]{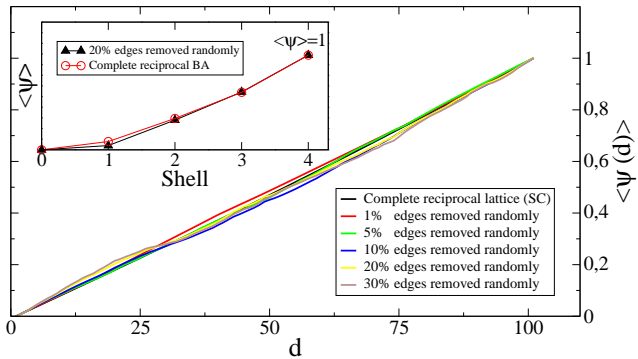}
    \caption{\label{fig1b} Plot of the electrostatic potential $\psi$, obeying the equation $\Delta \psi=0$ in a 2-d simple cube 
lattice and in a Barab\'asi-Albert network. In the simple cubic case the upper and lower layer are kept at the 
fixed value of $\psi =0$ and $\psi=1$ respectively. The average value increases from $0$ to $1$, even when the 
reciprocity in the links between pairs of nodes is broken up to 30\% of the completely reciprocal case. In the 
inset the plot of the same quantity for a Barab\'asi-Albert model where the same boundary conditions of $\psi=1$ 
and $\psi=0$ are applied to the leaves and to the core of the structure respectively. Here four shells are 
considered starting from a node in the core and the behaviour is the same even for a 20\% removal of reciprocal links}
\end{figure}

To clarify the role of $\Delta$ with respect to $\Delta_{C}$ we made some simple numerical tests. The traditional 
Laplacian operator describes a variety of different phenomena as for example the distribution of electrostatic potential in a dielectric when an electric field is applied to it (as for example by putting it into a capacitor). In this case the potential varies with the distance with the plates of the capacitor. We computed exactly this field in a series of lattices (both regular as the simple cube and some realizations of Barabasi-Albert network) starting from a completely reciprocal case and deleting randomly some of the connections. Generally, provided that the proportion of reciprocal links is above the percolation threshold for the lattice considered, the statistical behaviour of the directed Laplacian is the same of the non-directed one (See Figure \ref{fig1b}). 
\subsection{Expansion of the Wave Function}
The solution of the above eq. (\ref{8_ap}) is formally given by
\begin{equation}
\label{9_ap}
{\psi}=\left( -\Delta_{C}+V\right)^{-1}{F}  
\end{equation}
but for matrices of the size of the WWW ($\sim 10^{10}$ nodes) this solution cannot be used since it would be impossible to invert the matrix. We can make then use of matrix expansion through the following passages
\begin{equation}
 {\psi}=\left( I-V^{-1}\Delta_{C}\right)^{-1}V^{-1}{F} 
\end{equation}
the standard approach would be to use the following expansion
\begin{equation}
 \left( I-V^{-1}\Delta_{C}\right)^{-1}=\sum_{n=0}^{\infty}\left( V^{-1}\Delta_{C}\right)^{n}
\end{equation}
provided all the eigenvalues $\lambda_{h}$ of $\left( V^{-1}\Delta_{C}\right)$ are $| \lambda_{h}|<1$.\\
This expansion allows to invert only the diagonal matrix $V$ (that can be done easily by taking the inverse 
of the elements on the diagonal). While the physical meaning is clear in this form, it is mathematically more 
convenient  to simplify the expressions above by simplifying both the operator $\Delta_{C}$ and $V$ by putting eq. 
(\ref{9_ap}) in the simpler form
\begin{equation}
 {\psi}=\left( k^{O}-\alpha A^{T}\right)^{-1}F{'}=\left( I-\alpha B\right)^{-1}\left( k^{O}\right)^{-1}F{'}   
\end{equation}
where $F{'}=\alpha F$, $k^{O}$ is a matrix whose elements are all zero apart on the diagonal 
where they are given by the outdegree of 
vertices and $B=\left( k^{O}\right)^{-1}A^{T} $.\\
This is a form that closely resemble the original equation for PageRank, with the important caveat that we 
are now working with the wave function ${\psi}$. 
In this case the expansion:
\begin{equation}
 \left( I-\alpha B\right)^{-1}=\sum_{n=0}^{\infty}\left( \alpha B\right)^{n}
\end{equation}
does converge and we can calculate with the desired precision ${\psi}$ and so the associated Page Rank.
As a preliminary result this expansion outperformed by a factor three in speed the caclulation of the top 100 
of the .eu subset domain of the WWW.

\begin{acknowledgments}
G.C. acknowledges enlightening discussion with L. Pietronero. A.C. and N.P. thank G. Mula for 
fruitful suggestions. The activity of C.C. is funded by the ERC grant agreement 201766. A.C and N.P. 
acknowledge financial support by MIUR under project PON-CyberSar 2000-2006.
\end{acknowledgments}

\end{document}